\documentclass[10pt]{emulateapj}
\usepackage{graphicx}
\newcommand\bld[1]{\mbox{\boldmath $#1$}}

\newcommand\eps{{\epsilon}}

\newcommand{\pdv}[2]{\frac{\partial#1}{\partial#2}}
\newcommand{\dv}[2]{\frac{d#1}{d#2}}
\newcommand{\uv}[1]{\hat{\bld{#1}}}
\newcommand{\bnabla}{\bld{\nabla}}
\renewcommand{\bv}{\bld{v}}
\newcommand{\bxi}{\bld{\xi}}
\newcommand{\bB}{\bld{B}}
\newcommand{\bva}{\bv_A}
\newcommand{\bk}{\bld{k}}
\newcommand{\bO}{\bld{\Omega}}
\newcommand{\kdva}{\bv_A \cdot \bk}
\newcommand{\delr}{\delta \- \rho}
\newcommand{\delv}{\delta \bv}
\newcommand{\delb}{\delta \- \bB}
\newcommand{\bx}{v_{Ax}}
\newcommand{\delbx}{\delta v_{Ax}}
\newcommand{\delby}{\delta v_{Ay}}
\newcommand{\delbz}{\delta v_{Az}}
\newcommand{\wamp}[2]{\delta #1^{(#2)}}
\newcommand{\dwamp}[2]{\dot{\delta #1}^{(#2)}}
\newcommand{\be}{\begin{equation}}
\newcommand{\ee}{\end{equation}}
\newcommand{\bea}{\begin{eqnarray}}
\newcommand{\eea}{\end{eqnarray}}
\shortauthors{Johnson}
\shorttitle{}
\begin{document}

\title{Magnetohydrodynamic Shearing Waves}

\author{Bryan M. Johnson}

\affil{Astronomy Department,
601 Campbell Hall,
University of California at Berkeley,
Berkeley, CA 94720}

\begin{abstract}

I consider the nonaxisymmetric linear theory of a rotating, isothermal magnetohydrodynamic (MHD) shear flow. The analysis is performed in the shearing box, a local model of a thin disk, using a decomposition in terms of shearing waves, i.e., plane waves in a frame comoving with the shear. These waves do not have a definite frequency as in a normal mode decomposition, and numerical integration of a coupled set of amplitude equations is required to characterize their time dependence. Their generic time dependence, however, is oscillatory with slowly-varying frequency and amplitude, and one can construct accurate analytical solutions by applying the Wentzel-Kramers-Brillouin method to the full set of amplitude equations. The solutions have the following properties: 1) Their accuracy increases with wavenumber, so that most perturbations that fit within the disk are well-approximated as modes with time-dependent frequencies and amplitudes. 2) They can be broadly classed as incompressive and compressive perturbations, the former including the nonaxisymmetric extension of magnetorotationally unstable modes, and the latter being the extension of fast and slow modes to a differentially-rotating medium. 3) Wave action is conserved, implying that their energy varies with frequency. 4) Their shear stress is proportional to the slope of their frequency, so that they transport angular momentum outward (inward) when their frequency increases (decreases). The complete set of solutions constitutes a comprehensive linear test suite for numerical MHD algorithms that incorporate a background shear flow. I conclude with a brief discussion of possible astrophysical applications.

\end{abstract}


\section{Introduction}

The combined effects of magnetic fields and shear are important in such astrophysical systems as galactic disks \citep{shu74}, accretion disks \citep{bh91,bh98} and differentially-rotating stars \citep{spr02}. A simple theoretical model for understanding the local behavior of a magnetohydrodynamic (MHD) shear flow is the shearing box, developed originally for the study of gravitational spiral waves in galactic disks \citep{glb65,jt66} and used extensively in accretion disk theory. The model, which is local but fully nonlinear, is generated by an expansion of the equations of motion in the ratio of the disk scale height to the disk radius, and is therefore rigorously valid only for a thin disk. A great deal of insight into the behavior of astrophysical disks has been obtained through both analytical and numerical studies of the shearing box. This paper extends previous work on the linear theory of the MHD shearing box, with the primary goal of obtaining analytical solutions for use in testing numerical algorithms.

A linear analysis in the shearing box is complicated by the presence of the shear, the effect of which upon a plane wave of definite wavenumber is to rotate its wave crests into a trailing configuration with a primarily radial wavevector. The time dependence of these shearing waves, or ``shwaves'' \citep{jg05}, cannot in general be expressed in terms of a frequency eigenvalue as in a normal mode decomposition; one must solve instead a coupled set of differential equations. As a result, most linear analyses in the shearing box consider only axisymmetric modes, the spatial configurations of which remain stationary in the presence of shear. Characterization of the nonaxisymmetric shwaves generally requires numerical integration of the set of equations governing their time dependence (e.g., \citealt{bh92,cel97,fl97,bd06}), although approximate solutions have been obtained in some limiting cases. \cite{bh92} demonstrate the transient amplification of incompressive shwaves in the presence of a destabilizing field and show that extending the axisymmetric dispersion relation to general wavenumber provides a good approximation to the growth rate during the phase of nearly exponential amplification.\footnote{A similar analysis is performed by \cite{rg92} in the context of convective instability in disks.} \cite{ko00} analyze MHD shwaves in the limit of small shear to obtain what they refer to as ``coherent wavelet solutions''; these modal solutions are valid over a time scale that is short compared to the shearing time scale.

A glance at the numerical evolution of a set of shwaves, however, clearly indicates that their generic time dependence is oscillatory with a slowly varying frequency and amplitude, even in the presence of significant shear. Such a time dependence ought to be amenable to a Wentzel-Kramers-Brillouin (WKB) analysis; indeed, both \cite{bh92} and \cite{ko00} note that their dispersion relation represents the lowest order solution in a WKB expansion. The purpose of this paper is to generalize these analytical results into uniformly valid asymptotic solutions for both compressive and incompressive shwaves and for arbitrary values of the shear, and in particular to take the expansions to the next order and solve for the slowly-varying amplitudes.

The WKB method is typically applied to a second-order differential equation, but the method can be applied in a straightforward manner to a coupled set of differential equations of arbitrary order \citep{bre68}. With time as the independent variable, the WKB phase can be regarded as an integral over a local (``instantaneous'') frequency. The lowest order terms thus yield a time-dependent dispersion relation, a solution of which can be integrated to obtain the WKB phase (the eikonal equation). The next order terms yield a first-order differential equation for the slowly-varying amplitude of one of the dependent variables (the transport equation). The amplitudes of the other variables are then determined by the eigenvector components from the lowest order expansion. The problem is thus reduced to a set of decoupled equations for a slowly varying amplitude and phase which can be integrated by quadrature.\footnote{\cite{ft94,ft95} perform an analysis similar to what is described here, although their WKB expansion is performed in frequency space and they provide analytical expressions only for strongly leading and trailing shwaves.} Alternatively, the overall amplitude can be determined from the constraint that wave action is conserved, a general result for oscillatory solutions derived from a Lagrangian. 

I begin in \S \ref{section_be} with an overview of the local equations for an isothermal MHD shear flow in their general and linearized forms, as well as the decomposition of the latter in terms of shwaves. \S\ref{section_is} and \S\ref{section_cs} describe the leading order solutions for incompressive and compressive shwaves, respectively, and \S\ref{SEAMT} discusses their energy and angular momentum transport properties. I summarize and discuss applications in \S \ref{section_summ}.

\section{Basic Equations}\label{section_be}

As discussed in the Introduction, the shearing box is generated by an expansion of the equations of motion in the ratio of the disk scale height to the disk radius. The resulting model is a local Cartesian frame in which the orbital velocity varies linearly with radius. The shearing box equations in isothermal MHD are

\be\label{BE1}
\pdv{\rho}{t} + \bnabla \cdot \left(\rho \bv \right) = 0,
\ee

\bea\label{BE2}
\pdv{\bv}{t} + \bv\cdot \bnabla \bv + c_s^2\frac{\bnabla\rho}{\rho} + \frac{\bnabla B^2}{8\pi \rho} 
- \frac{\bB\cdot \bnabla\bB}{4\pi \rho}  \nonumber \\ \;  + 2 \bO \times \bv - 2q\Omega^2 x \uv{x} = 0,
\eea

\be\label{BE3}
\pdv{\bB}{t} - \bnabla \times \left(\bv \times \bB \right) = 0,
\ee
and
\be\label{BE4}
\bnabla \cdot \bB = 0,
\ee
where $\rho$ is the volume density, $\bv$ is the bulk fluid velocity, $c_s$ is the isothermal sound speed, $\bB$ is the magnetic field, $\bO = \Omega \uv{z}$ is the rotation vector and 
\be
q \equiv -\frac{1}{2}\dv{\ln \Omega^2}{\ln r}
\ee
is the shear parameter. The last term in equation (\ref{BE2}) is the radial tidal force and the penultimate term is the local Coriolis force.\footnote{There is an additional vertical tidal term $\Omega^2 z \uv{z}$ which I have ignored for simplicity; this is equivalent to considering scales much smaller than a disk scale height.} The coordinates $x, y$ and $z$ correspond to the radial, azimuthal and vertical directions, respectively. 

The equilibrium state is one of constant density $\rho_0$, spatially constant magnetic field $\bB_0$ and orbital velocity
\be
\bv_0 = -q\Omega x \, \uv{y}.
\ee
In linearizing about this equilibrium state, it is convenient to work in units in which the mean density $\rho_0$ is unity, so that the magnetic field has units of velocity. In these units,
\be\label{DRHO}
\rho = 1 + \delr,
\ee
\be
\bv = \bv_0 + \delv,
\ee
\be
\frac{\bB}{\sqrt{4\pi \rho}} = \bv_A + \delb,
\ee
where $\bv_A \equiv \bB_0/\sqrt{4\pi}$ is the Alfv\`en velocity.

In the absence of fixed boundaries, the natural basis for local perturbations in the shearing box is a decomposition in terms of plane waves in shearing coordinates, $(x^\prime, y^\prime, z^\prime) \equiv (x, y - v_0 t, z)$:
\be\label{PERT}
\delta(t) \exp\left(i \bk \cdot \bld{x}\right) = \delta(t) \exp\left(i k_x[0] x + i k_y [y - v_0 t] + i k_z z\right).
\ee
A non-shearing observer sees these as plane waves with a time-dependent radial wavenumber:
\be
k_x(t) \equiv k_x(0) + q\Omega k_y t.
\ee 
As a result of the time dependence of their radial gradients, nonaxisymmetric shwaves do not have a definite frequency as in a normal mode decomposition. One can readily verify from the induction equation (\ref{BE3}) that 
\be\label{VA}
v_{Ay}(t) = v_{Ay}(0) - q\Omega v_{Ax} t,
\ee
so that $v_A$ also varies linearly with time in the presence of a radial field. 

Expanding equations (\ref{BE1})-(\ref{BE4}) to linear order in the perturbation variables defined in equations (\ref{DRHO})-(\ref{PERT}) results in
\be\label{L1}
\dot{\delr} + i\bk \cdot \delv = 0,
\ee
\vspace{-0.2in}
\bea\label{L2}
\dot{\delv} + ic_s^2 \delr \, \bk  + i (\bv_A \cdot \delb) \bk
- i (\kdva) \delb \nonumber \\ \; - 2\Omega \delta v_y \uv{x} + (2 - q)\Omega \delta v_x \uv{y} = 0,
\eea
\be\label{L3}
\dot{\delb} - i(\kdva) \delv + i (\bk \cdot \delv) \bv_A 
+ q\Omega \delbx \uv{y} = 0,
\ee
and
\be\label{L4}
\bk \cdot \delb = 0,
\ee
where overdots denote time derivatives and where the common factor $\exp\left(i\bk \cdot \bld{x}\right)$ has been removed. Equations (\ref{L1})-(\ref{L4}) are solved in various limits in the following sections.

\section{Incompressive Shwaves}\label{section_is}

\subsection{Equations}

By definition, incompressive shwaves satisfy
\be\label{KDV0}
\bk \cdot \delv \approx 0
\ee
throughout their evolution. The appropriate limit for extracting them is therefore the Boussinesq approximation, in which $\partial_t \ll c_s k$. This implies both a subthermal field ($\beta \equiv c_s^2/v_A^2 \gg 1$) and wavelengths that are short compared to the disk scale height ($c_s k/\Omega \gg 1$). The behavior of these transverse shwaves can be isolated by taking the cross product of equation (\ref{L2}) with $\bk$ twice.\footnote{This was pointed out to me by Yoram Lithwick.} Combining the resulting equation with the induction equation and the divergence-free constraints on the velocity and magnetic field perturbations, one obtains the following reduced set of linear equations:
\be\label{Li1}
\dot{\delta \eta} - i (\kdva) k^2 \delta v_{Ax}  
+ 2\Omega k_z \delta \omega_x = 0,
\ee
\be\label{Li2}
\dot{\delta \omega_x} - i (\kdva) \delta J_x - (2 - q) \Omega k_z \frac{\delta \eta}{k^2} = 0,
\ee
\be\label{Li3}
\dot{\delta v_{Ax}} - i (\kdva) \frac{\delta \eta}{k^2} = 0,
\ee
\be\label{Li4}
\dot{\delta J_x} - i (\kdva) \delta \omega_x - q\Omega k_z \delta v_{Ax} = 0,
\ee
where
\be
\delta \eta \equiv k^2 \delta v_x
\ee
is proportional in two-dimensions to the $z$-component of the perturbed vorticity,
\be
\delta \omega_x \equiv k_y \delta v_z - k_z \delta v_y
\ee
is the $x$-component of the perturbed vorticity, and
\be
\delta J_x \equiv k_y \delta v_{Az} - k_z \delta v_{Ay}
\ee
is proportional to the $x$-component of the perturbed current. The self-consistent density perturbation is obtained by taking the dot product of equation (\ref{L2}) with $\bk$:
\be\label{DRHOI}
\delr = -\frac{\bv_A \cdot \delb}{c_s^2} - i \frac{2\Omega}{c_s k}\left(\frac{k_x}{k} \frac{\delta v_y}{c_s} + [q-1]\frac{k_y}{k} \frac{\delta v_x}{c_s}\right).
\ee
In deriving the above equations, it is important to notice that
\be
\bk \cdot \dot{\delv} = \dv{}{t}\left(\bk \cdot \delv\right) - \dv{\bk}{t} \cdot \delv \approx -q\Omega k_y \delta v_x
\ee
due to the time dependence of $k$.

\subsection{Solution}\label{SIS}

Without loss of generality, one can express each perturbation variable in a WKB form:
\be\label{WKB}
\delta(t) \equiv \sum_{n = 0}^{\infty} \delta^{(n)}(t) \exp\left(-i\int \omega(t) \, dt\right),
\ee
where the WKB phase has been expressed in terms of an integral over a slowly varying frequency. For the incompressive shwaves, it is not obvious a priori what small parameter to use in the expansion, but clearly a valid expansion requires $\delta^{(n+1)} \ll \delta^{(n)}$, as well as $\partial_t(\ln \omega) \ll \omega$ (the adiabatic approximation).\footnote{An additional requirement is that $\partial_t\ln \delta^{(n)} \ll \omega$, but since $\delta^{(n)}$ is a function of $\omega$, a slowly-varying frequency implies a slowly-varying amplitude.}

Expressing each perturbation variable in the form (\ref{WKB}), the lowest-order terms from equations (\ref{Li1})-(\ref{Li4}) result in a dispersion relation with the same form as the axisymmetric dispersion relation \citep{bh91,bh92}:
\be\label{DRi}
\tilde{\omega}^4 - \frac{\kappa^2 k_z^2}{k^2} \, \tilde{\omega}^2 - \frac{4\Omega^2 k_z^2}{k^2} \, (\kdva)^2 = 0,
\ee
where $\kappa^2 \equiv 2(2 - q)\Omega^2$ is the epicyclic frequency and
\be\label{TOM}
\tilde{\omega}^2 \equiv \omega^2 - (\kdva)^2.
\ee
The differences here are that $k$ is the full three-dimensional wavenumber so that $\omega$ depends upon time via $k_x(t)$.\footnote{Notice that while $\bv_A$ and $\bk$ both vary linearly with time, $\kdva$ is a constant.} The eigenvector components can be expressed as
\be\label{EVi1}
\delv = \frac{\omega}{\kdva} \, \bk \times \left(\bk \times \uv{x} + i\frac{\tilde{\omega}^2 k^2}{2 \Omega k_z \omega} \uv{x}\right) \frac{\delbx}{k_y^2 + k_z^2}
\ee
and
\be\label{EVi2}
\delb = -\bk \times \left(\bk \times \uv{x} + i\frac{2\Omega k_z \omega}{\tilde{\omega}^2} \uv{x}\right)\frac{\delbx}{k_y^2 + k_z^2}.
\ee

As demonstrated explicitly in Appendix~\ref{FOIS}, the next order in the WKB expansion yields the time dependence of the slowly-varying amplitude. An alternative approach is to compute the amplitude from the conservation of wave action. It is well known that slowly-varying wavetrains in non-dissipative, continuous systems obey the following conservation law to all orders in a WKB expansion \citep{whit65,bg68,whit70}:
\be\label{WAC}
\pdv{}{t}\left(\frac{{\cal E}}{\omega}\right) + \bnabla \cdot \left(\frac{{\cal E}}{\omega} \, \bv_g \right) = 0,
\ee
where ${\cal E}$ is a phase-averaged perturbation energy, $\omega$ is the frequency as measured by an observer moving with the local velocity of the medium, $\bv_g \equiv \bnabla_{\bk} \omega$ is the group velocity, and ${\cal E}/\omega$ is referred to as the wave action density. Since the phase-averaged energy of a shwave is spatially constant, expression (\ref{WAC}) in this context reduces to
\be\label{WAC2}
\pdv{}{t}\left(\frac{{\cal E}}{\omega}\right) = 0.
\ee

The perturbation energy appearing in the adiabatic invariant ${\cal E}/\omega$ is
\be\label{Edef}
{\cal E} \equiv \omega \pdv{{\cal L}}{\omega} - {\cal L},
\ee
where ${\cal L}$ is a phase-averaged perturbation Lagrangian.\footnote{This is generally not equivalent to an expansion of the energy expressed in terms of Eulerian perturbations.} The energy as defined in expression (\ref{Edef}) is derived in Appendix~\ref{section_se}; in the incompressible limit, expression (\ref{SE}) for the average shwave energy reduces to
\be\label{PEi}
{\cal E} = \frac{1}{4}\left(\omega^2\left| \xi \right|^2 - 2q\Omega^2\left| \xi_x \right|^2 + \left|\delta v_A \right|^2\right),
\ee
where $|\xi|^2 \equiv \bxi \cdot \bxi^*$ is the magnitude of the Lagrangian fluid displacement $\bxi$ and $|\delta v_A |^2 \equiv \delb \cdot \delb^*$. The relation between $\bxi$ and the Eulerian velocity perturbation in a differentially-rotating medium is given by 
\be\label{Disp}
\delv = \dot{\bxi} + q\Omega \xi_x \uv{y}.
\ee

Computing ${\cal E}$ using the eigenvector relations (\ref{EVi1}) and (\ref{EVi2}) subject to the constraint ${\cal E} \propto \omega$ from (\ref{WAC2}) gives\footnote{This result also requires taking the limit $k_z \gg k_y$; see \S\ref{VEI}.}
\be\label{DBXamp}
\delbx \propto \sqrt{\frac{\tilde{\omega}^2}{\omega\left(2\tilde{\omega}^2 k^2 -  \kappa^2 k_z^2\right)}},
\ee
so that to within a multiplicative constant the amplitude of the incompressive shwaves is given by
\be\label{VI}
\delv = \tilde{\omega} \sqrt{\frac{\omega}{2\tilde{\omega}^2 k^2 -  \kappa^2 k_z^2}} \, \bk \times \left(\bk \times \uv{x} + i\frac{\tilde{\omega}^2 k^2}{2 \Omega k_z \omega}\uv{x}\right)
\ee
and
\be\label{BI}
\delb = -\tilde{\omega}\, \frac{\kdva}{\omega}\sqrt{\frac{\omega}{2\tilde{\omega}^2 k^2 -  \kappa^2 k_z^2}} \, \bk \times \left(\bk \times \uv{x} + i\frac{2\Omega k_z \omega}{\tilde{\omega}^2} \uv{x}\right).
\ee
Notice that these are nonlinear solutions to the basic equations, a general result for a single incompressive plane wave perturbation \citep{gx94}.

\subsection{Validity of Expansion}\label{VEI}

For wavenumbers with $\kdva \gg \Omega$, both branches of the dispersion relation (\ref{DRi}) reduce to
\be
\omega^2 = \left(\kdva\right)^2.
\ee
Notice that these short-wavelength modes do not have $\bk \parallel \bv_A$ (i.e., they are not shear Alfv\`enic), since $\theta(t) \rightarrow \pi/2$ due to the shear, where $\theta$ is the angle between $\bv_A$ and $\bk$.

For wavenumbers with $\kdva \sim \Omega$, an accurate expansion generally requires $k_z \gg k_y$. One way to see this is to rewrite equations (\ref{Li1})-(\ref{Li4}) in terms of $\delta v_x$, which introduces an additional term $(2q\Omega k_x k_y)\delta v_x$ into equation (\ref{Li1}). The resulting dispersion relation matches (\ref{DRi}) in the limit $k_z \gg k_y$ (for $q \sim 1$).

For wavenumbers satisfying
\be\label{ISsplit}
\frac{16\Omega^2 k^2 \left(\kdva\right)^2}{\kappa^4 k_z^2} \ll 1,
\ee
the two branches of the dispersion relation (\ref{DRi}) reduce to
\be\label{DRi2}
\omega_+^2 = \kappa^2\frac{k_z^2}{k^2}
\ee
and
\be
\omega_-^2 = \left(\kdva\right)^2\left(1 - \frac{4\Omega^2}{\kappa^2}\right).
\ee
Notice that $\omega_-^2 < 0$ for $q < 0$, reflecting the unstable nature of this branch of the dispersion relation \citep{bh92}. The validity of expression (\ref{ISsplit}) generally requires $\kdva \ll \Omega$; this implies either $\kdva \approx 0$ or, since the wavenumbers must also satisfy the Boussinesq approximation, $\Omega (v_A/c_s) \ll v_A k \ll \Omega$, i.e., $\beta^{1/2} \gg c_s k/\Omega \gg 1$. In the limit of a very weak field or for wavevectors oriented nearly perpendicular to $\bv_A$, there is thus a range of wavenumbers satisfying (\ref{ISsplit}) that increases with decreasing field strength. Eventually (\ref{ISsplit}) breaks down, however, due to the increase of $k$ with time, and the frequency will asymptote to $\kdva$.

Equation (\ref{DRi2}) satisfies the adiabatic approximation for 
\be\label{EIKi}
\left|\frac{\partial_t (\ln\omega_+)}{\omega_+}\right| =  \frac{q\Omega k_y |k_x|}{\kappa k_z k} \ll 1.
\ee
This ratio is a (constant) maximum at large $|k_x(t)|$, and therefore a uniformly valid expansion requires 
\be\label{EIKi2}
\frac{k_z}{k_y} \gg \frac{q\Omega}{\kappa}.
\ee

\begin{figure}
\plotone{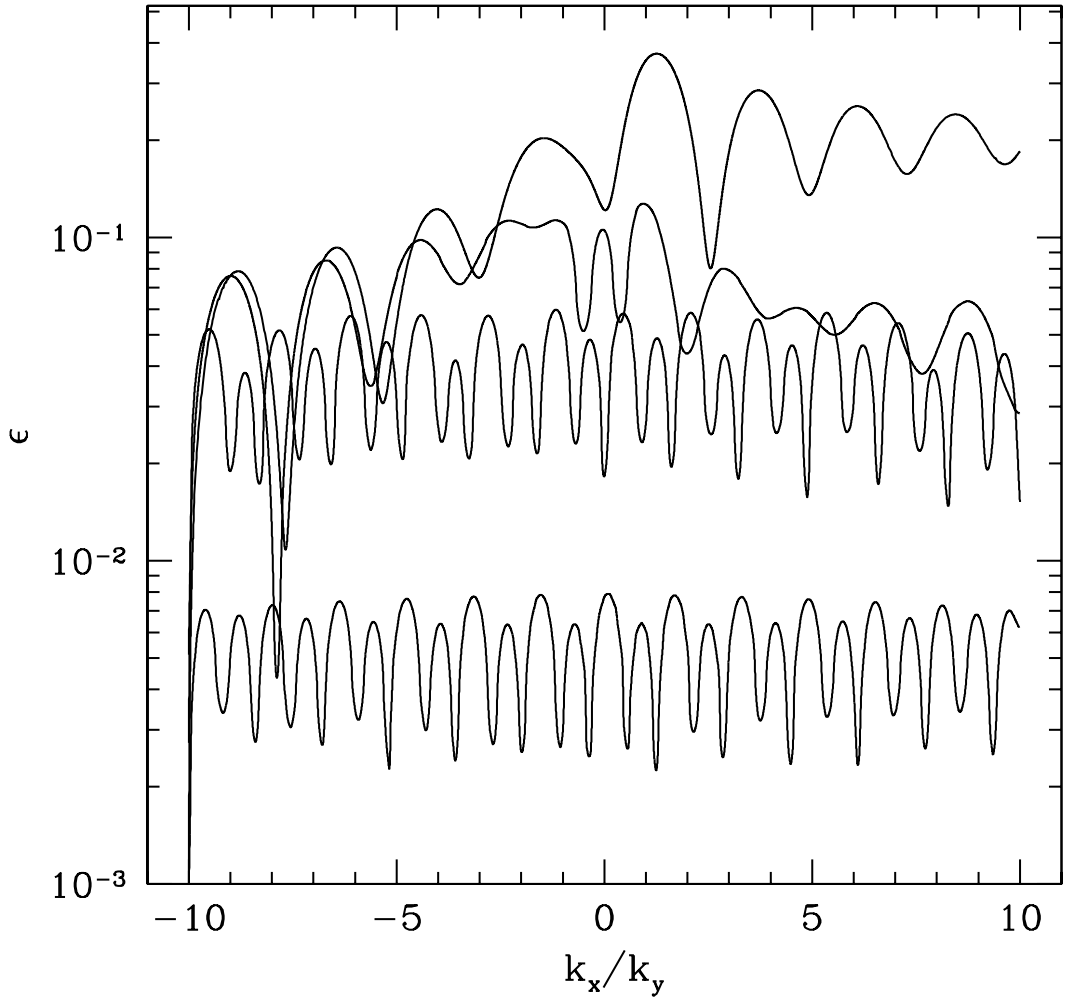}
\caption{
Error in the energy of an incompressive shwave corresponding to the positive branch of the dispersion relation (\ref{DRi}) for a series of numerical integrations with $c_s k_z/(2\pi\Omega) = 10$ and $k_y/k_z$ varying from $1$ (top) to $10^{-4}$ (bottom) by factors of $10$. The data have been smoothed for readability. The shear parameter $q = 3/2$, corresponding to a Keplerian rotation profile.
}
\label{f1}
\end{figure}

These solutions have been verified with a numerical integration of equations (\ref{L1})-(\ref{L4}), with excellent agreement in general for wavenumbers satisfying $k_z \gg k_y$. Figure~\ref{f1} shows the error in the solution corresponding to the positive branch of the dispersion relation (\ref{DRi}) for a series of numerical integrations with varying $k_z/k_y$, $k_x(0) = -10 k_y$ and $q = 3/2$ (Keplerian rotation). The vertical wavenumber is fixed at $c_s k_z/(2\pi\Omega) = 10$ and the mean field is $v_{Az} = 2\Omega/k_z$. The error is defined to be
\be
\eps \equiv \sqrt{\frac{|{\cal E}_n - {\cal E}_a|}{{\cal E}_{a}}},
\ee
where ${\cal E}_n$ and ${\cal E}_a$ are the numerical and analytical values of expression (\ref{PEi}). A numerical check of the accuracy of the solutions in the limit $k_y \gg k_z$ indicates that the errors remain small ($\lesssim 30\%$).

\begin{figure}
\plotone{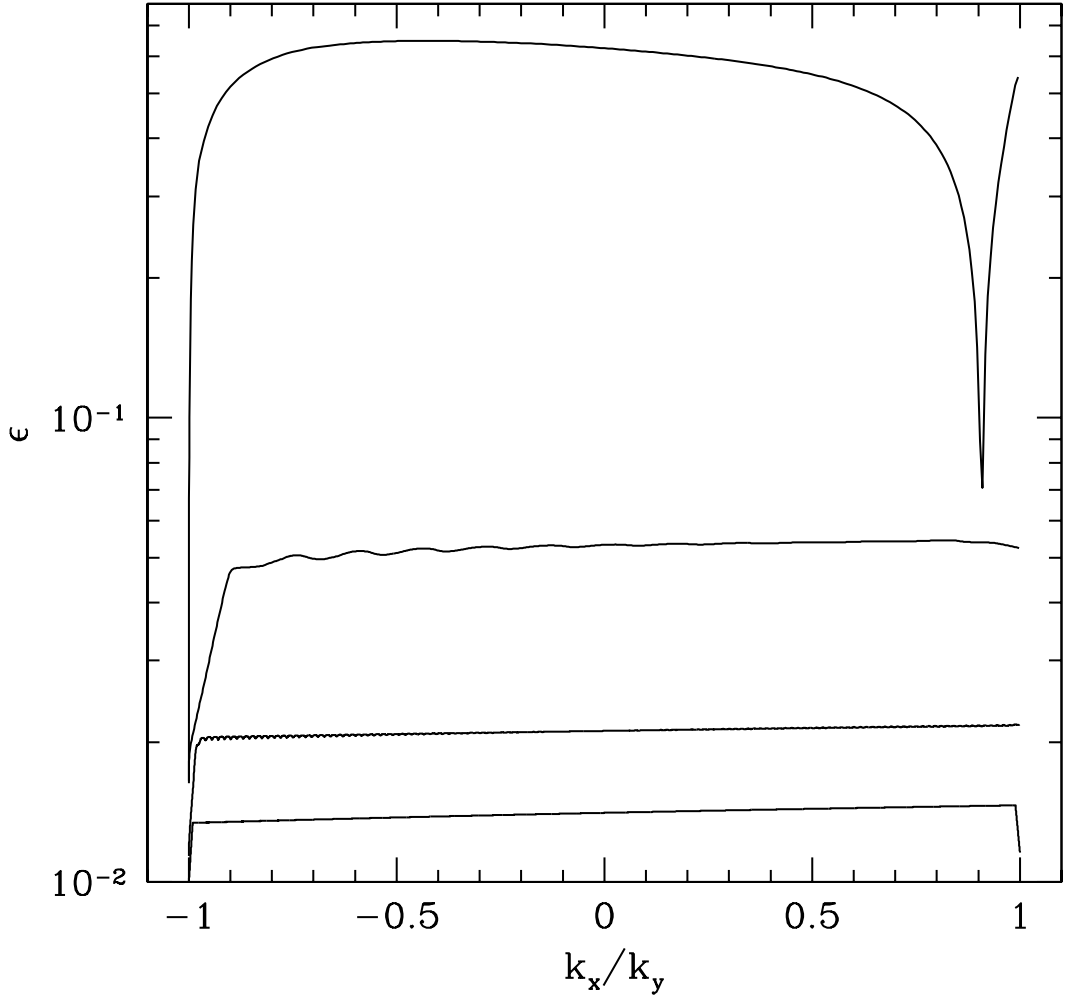}
\caption{
Error in the energy of an incompressive shwave corresponding to the negative branch of the dispersion relation (\ref{DRi}) for a series of numerical integrations with $q = 3/2$, $c_s k_y/(2\pi\Omega) = 1$ and $k_z/k_y = 3$ (top), $10$, $30$, $100$ and $300$ (bottom). The data have been smoothed for readability.
}
\label{f2}
\end{figure}

Figure~\ref{f2} shows the error in the solution corresponding to the negative branch of the dispersion relation (\ref{DRi}) for a series of numerical integrations with varying $k_z/k_y$ and $k_x(0) = -k_y$. The azimuthal wavenumber is fixed at $c_s k_y/(2\pi\Omega) = 1$ and the mean field is $v_{Az} = \sqrt{15/16}\,\Omega/k_z$, corresponding to the most unstable wavenumber in axisymmetry.

Notice that for wavenumbers satisfying
\be\label{KZCRIT}
\frac{k_z^2}{k_y^2}  > \frac{\left(\kdva\right)^2}{2q\Omega^2 - \left(\kdva\right)^2},
\ee
the lower branch of the dispersion relation (\ref{DRi}) goes to zero at
\be\label{KXCRIT}
k_x(t) = \pm k_y \sqrt{\frac{k_z^2}{k_y^2}\left(\frac{2q\Omega^2}{\left[\kdva\right]^2} -1\right) - 1}.
\ee
While the solution is accurate in both the oscillatory and exponential regimes, it breaks down as $\omega \rightarrow 0$, and a uniformly valid expansion would require a more careful treatment of the solution near these turning points.

A numerical check of the incompressive solutions at varying values of $c_s k/\Omega$ indicates that compressibility effects are generally negligible for $c_s k/(2\pi\Omega) \gtrsim 3$.

\subsection{Weak Field Limit}

For $\kdva \ll \Omega$ and $k_z/k_y \lesssim q\Omega/\kappa$, condition (\ref{EIKi}) is violated and one must revisit the Boussinesq equations (\ref{Li1})-(\ref{Li4}) in that limit. For general $k_z/k_y$ and $\kdva \ll \Omega$, these equations can be combined into a second-order differential equation,
\be
k^2 \ddot{\delbx} + 2q\Omega k_x k_y \dot{\delbx} + \kappa^2 k_z^2 \delbx = 0,
\ee
which can be solved in terms of hypergeometric functions:
\bea\label{Bxi}
\delbx = c_1\, F\left(\frac{1 - \alpha}{4},\frac{1 + \alpha}{4},\frac{1}{2},-\frac{k_x^2}{k_y^2 + k_z^2}\right) \nonumber \\ \; + c_2 \, k_x \, F\left(\frac{3 - \alpha}{4},\frac{3 + \alpha}{4},\frac{3}{2},-\frac{k_x^2}{k_y^2 + k_z^2}\right),
\eea
where
\be
\alpha \equiv \sqrt{1 - \frac{4\kappa^2k_z^2}{q^2\Omega^2k_y^2}}
\ee
and $c_1$ and $c_2$ are integration constants that depend on the initial conditions. The other perturbation variables are given by\footnote{The hypergeometric functions have simple integration and differentiation relations (e.g., \citealt{as72}) and can be implemented for small argument by the GNU Scientific Library. A Mathematica script that generates the solutions described here is available upon request.}
\be
\delta J_x = 2\Omega k_z \int \delbx \, dt,
\ee
\be\label{etai}
\delta \eta = -i\frac{k^2}{\kdva}\dot{\delbx},
\ee
and
 \be\label{omxi}
 \delta \omega_x = -i\frac{(2-q)\Omega k_z}{\kdva} \delbx,
 \ee
 with
 \be\label{vyi}
 \left(\delta v_y,\delby\right) = -\frac{k_z\left(\delta \omega_x, \delta J_x\right) + k_x k_y\left(\delta v_x, \delbx\right)}{k_y^2 + k_z^2}
 \ee
 and
 \be\label{vzi}
 \left(\delta v_z,\delbz\right) = \frac{k_y\left(\delta \omega_x, \delta J_x\right) - k_x k_z\left(\delta v_x, \delbx\right)}{k_y^2 + k_z^2}.
 \ee

\subsection{Hydrodynamic Limit}

In the absence of magnetic fields, the amplitude of the incompressive shwaves is given by
\be\label{VHYDRO}
\delv = \sqrt{\omega} \, \bk \times \left(\uv{k} \times \uv{x} + i\frac{\kappa}{2 \Omega} \uv{x}\right)
\ee
and
\be
\delr = \frac{\sqrt{\omega}}{c_s k}\left(\frac{\kappa k_z k_x}{c_s k} - i \frac{2\Omega k_y}{c_s} \left[1 - q\frac{k_y^2 + k_z^2}{k^2}\right]\right),
\ee
with $\omega = \kappa k_z/k$.

These are the WKB solutions analyzed by \cite{amn05} and \cite{bh06} in the context of enhanced angular momentum transport in unmagnetized disks.\footnote{Notice that the energy amplification of these solutions \citep{amn05} is due to the time dependence of $\omega$ and the conservation of wave action; see \S\ref{SEAMT}.} They are uniformly valid for all wavenumbers satisfying condition (\ref{EIKi2}). For general wavenumber, the solutions can again be expressed in terms of hypergeometric functions \citep{bh06}.

\section{Compressive Shwaves}\label{section_cs}

\subsection{Equations}

In the absence of rotation, the compressive modes are the fast and slow waves. They have $\partial_t \gtrsim c_s k$ and a nonzero velocity component along their propagation direction. It is therefore natural to work with the linear equations in the following form:
\bea\label{Lc1}
\bk \cdot \dot{\delta \bv} + i k^2\left(c_s^2 \delta \rho + \bv_A \cdot \delta \bva \right) \nonumber \\ \; + (2 - q) \Omega k_y \delta v_x - 2 \Omega k_x \delta v_y = 0,
\eea
\vspace{-0.2in}
\bea\label{Lc2}
\bv_A \cdot \dot{\delta \bv} + i \left(\kdva\right)c_s^2 \delta \rho  \nonumber \\ \; + (2 - q) \Omega v_{Ay} \delta v_x - 2 \Omega v_{Ax} \delta v_y = 0,
\eea 
and
\bea\label{Lc3}
\bv_A \cdot \dot{\delta \bva} - i\left(\kdva\right) \bv_A \cdot \delta \bv + i v_A^2 \bk \cdot \delta \bv \nonumber \\ \; + q\Omega v_{Ay} \delbx = 0,
\eea
along with the continuity equation (\ref{L1}).

\subsection{Solution}\label{SCS}

Establishing a uniformly valid expansion for compressive shwaves requires taking the limit $H k \gg 1$, where $H \equiv \sqrt{c_s^2 + v_A^2}/\Omega$; these short wavelengths are dynamically unaffected by the rotation. The lowest-order terms in the expansion yield the dispersion relation
\be\label{DRc}
\omega^4 - \left(c_s^2 + v_A^2\right) k^2 \omega^2 + c_s^2 k^2 \left(\kdva\right)^2 = 0,
\ee
with eigenvector components
\be\label{EVc1}
\delv =\frac{\omega}{\tilde{\omega}^2}\left(\frac{\omega^2}{k^2} \bk - \kdva \, \bv_A \right) \delr
\ee
and
\be\label{EVc2}
\delb = \frac{\omega^2}{\tilde{\omega}^2}  \left(\bv_A - \frac{\kdva}{k^2} \, \bk \right) \delr.
\ee

For short-wavelength compressive shwaves, expression (\ref{SE}) for the average shwave energy reduces to
\be\label{PEc}
{\cal E} = \frac{1}{4}\left(\left| \delta v \right|^2 + c_s^2 \left| \delr \right|^2 + \left|\delta v_A \right|^2\right),
\ee
where $\left| \delta v \right|^2 \equiv \delv \cdot \delv^*$ ($=\omega^2 \left| \xi \right|^2$ to leading order in $[Hk]^{-1}$) and $|\delr |^2 \equiv \delr \, \delr^*$. Computing ${\cal E}$ using the eigenvector relations (\ref{EVc1}) and (\ref{EVc2}), subject to the constraint ${\cal E} \propto \omega$, gives
\be\label{DRamp}
\delr
\propto \tilde{\omega}k\sqrt{\frac{\omega}{\omega^4 - c_s^2 k^2 \left(\kdva\right)^2}}.
\ee
As demonstrated explicitly in Appendix~\ref{FOCS}, the same result is obtained by going to the next order in the WKB expansion. To within a multiplicative constant, then, the amplitude of the compressive shwaves is given by
\be
\delv = \frac{\omega k}{\tilde{\omega}}\sqrt{\frac{\omega}{\omega^4 - \left(\kdva\right)^2 c_s^2 k^2}}\left(\frac{\omega^2}{k^2} \bk - \kdva \, \bv_A \right),
\ee
\be
\delb = \frac{\omega^2 k}{\tilde{\omega}}\sqrt{\frac{\omega}{\omega^4 - \left(\kdva\right)^2 c_s^2 k^2}}\left(\bv_A - \frac{\kdva}{k^2} \, \bk \right),
\ee
and
\be
\delr = \frac{\tilde{\omega} k}{c_s}\sqrt{\frac{\omega}{\omega^4 - \left(\kdva\right)^2 c_s^2 k^2}},
\ee

\subsection{Validity of Expansion}

In the limit
\be
\frac{\left(\kdva\right)^2 4\beta}{v_A^2 k^2\left(1 + \beta\right)^2} \ll 1,
\ee
which is valid unless $\beta \sim 1$, the frequencies of the fast and slow shwaves reduce to
\be
\omega_+^2 = \left(c_s^2 + v_A^2\right) k^2 = c_s^2 k^2 \left(1 + \beta\right)
\ee
(magnetosonic shwaves) and
\be
\omega_-^2 = \frac{c_s^2 \left(\kdva\right)^2}{c_s^2 + v_A^2} = \frac{\beta}{1 + \beta}\left(\kdva\right)^2,
\ee
respectively.

For a mean field that is constant with time ($v_{Ax} = 0$), the fast shwaves satisfy the adiabatic approximation for 
\be\label{EIKf}
\left|\frac{\partial_t (\ln\omega_+)}{\omega_+}\right| =  \frac{q k_y}{H}\frac{\left|k_x\right|}{k^3} \ll 1.
\ee
This has a maximum value at $k_x^2 = (k_y^2 + k_z^2)/2$, so that a uniformly valid expansion requires
\be
\left|\frac{\partial_t (\ln\omega_+)}{\omega_+}\right|_{max} = \frac{2q k_y}{3 H \sqrt{3}(k_y^2 + k_z^2)}  \sim \frac{q}{H k} \ll 1.
\ee

For $\beta \gg 1$, the slow shwaves have $\partial_t \sim v_A k \ll c_s k$ and the basic assumption for compressive shwaves is no longer valid. The solution for the slow shwaves is therefore accurate only for $\beta \lesssim 1$. Even in this regime the solution breaks down asymptotically in the presence of a time-dependent magnetic field, since in that case $\omega_- \propto \sqrt{\beta}$ decreases with time; the breakdown occurs when
\be
\frac{\omega_-}{c_s k} \sim \frac{\kdva}{v_A k} = \cos \theta \ll 1,
\ee
i.e., when $\bv_A$ and $\bk$ are nearly perpendicular.

\begin{figure}
\plotone{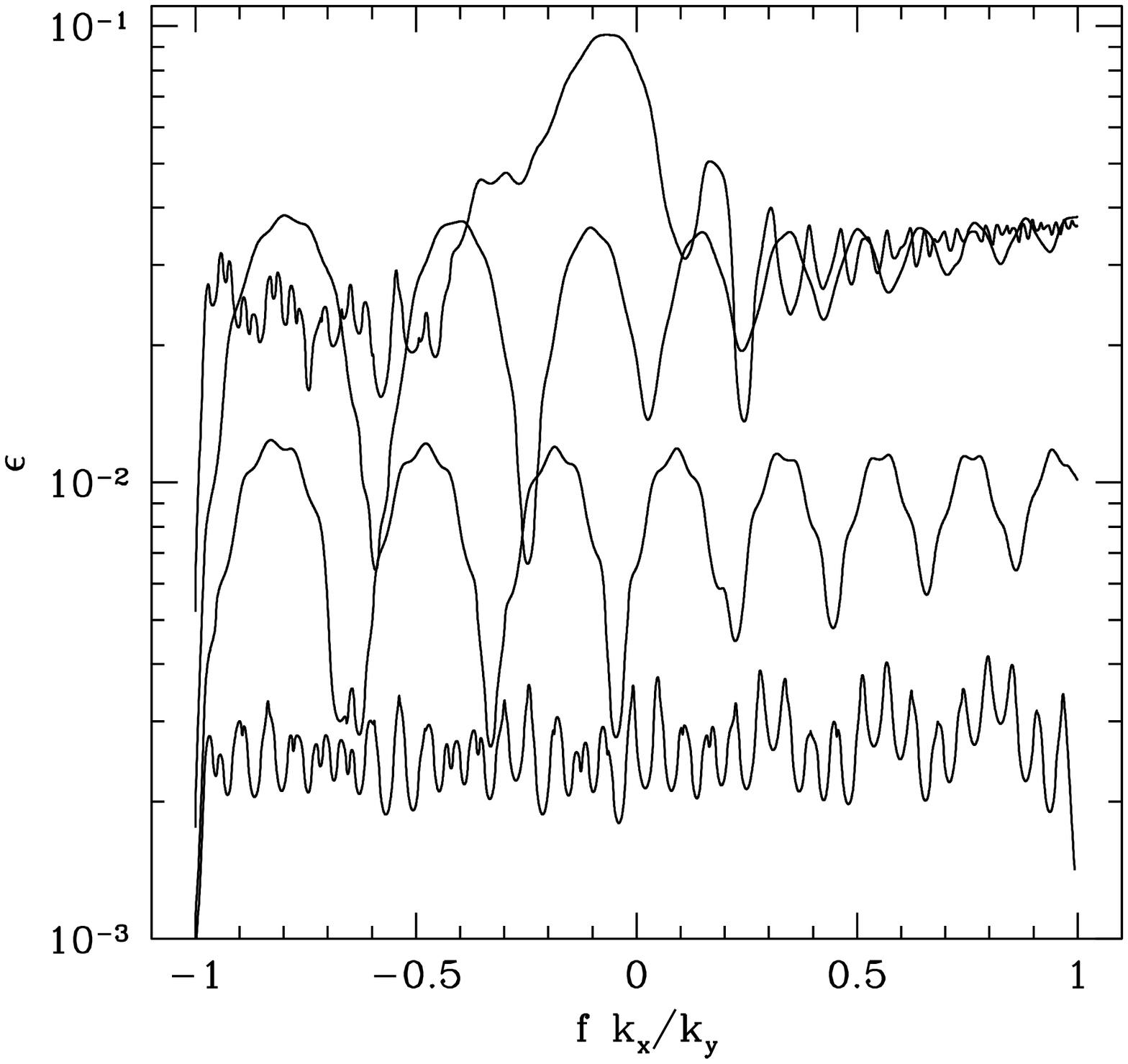}
\caption{
Error in the energy of a fast shwave for a series of numerical integrations with $q = 3/2$ and $c_s k_y/(2\pi\Omega) = c_s k_z/(2\pi\Omega) = 1$ (top), $3$, $10$, $30$ and $100$ (bottom). The scale factor $f$ is defined in expression (\ref{SCALE}). The data have been smoothed for readability.
}
\label{f3}
\end{figure}

These solutions have been verified with a numerical integration of equations (\ref{L1})-(\ref{L4}), with excellent agreement in general for wavenumbers satisfying $Hk/(2\pi) \gg 1$. Figure~\ref{f3} shows the error in the energy of a fast shwave for a series of numerical integrations with varying $k_z = k_y$ and $q = 3/2$. The initial mean field is $v_{Ax} = -c_s$, $v_{Ay} = 3c_s$ and $v_{Az} = c_s$. A consistent comparison of the errors in this case is difficult since the frequency of the compressive shwaves increases with $k$ and therefore the requirements for an accurate numerical integration become more stringent. To make a fair comparison, the integration time is scaled with $k_z$ and $k_y$ to give approximately the same number of oscillations during each integration. For $k_z = k_y$, the scale factor $f$ in Figure~\ref{f3} is given by
\be\label{SCALE}
f \equiv \left(\sqrt{1 + 99\left[\frac{2\pi \Omega}{c_s k_y}\right]^2} - 1\right)^{1/2}.
\ee

\begin{figure}
\plotone{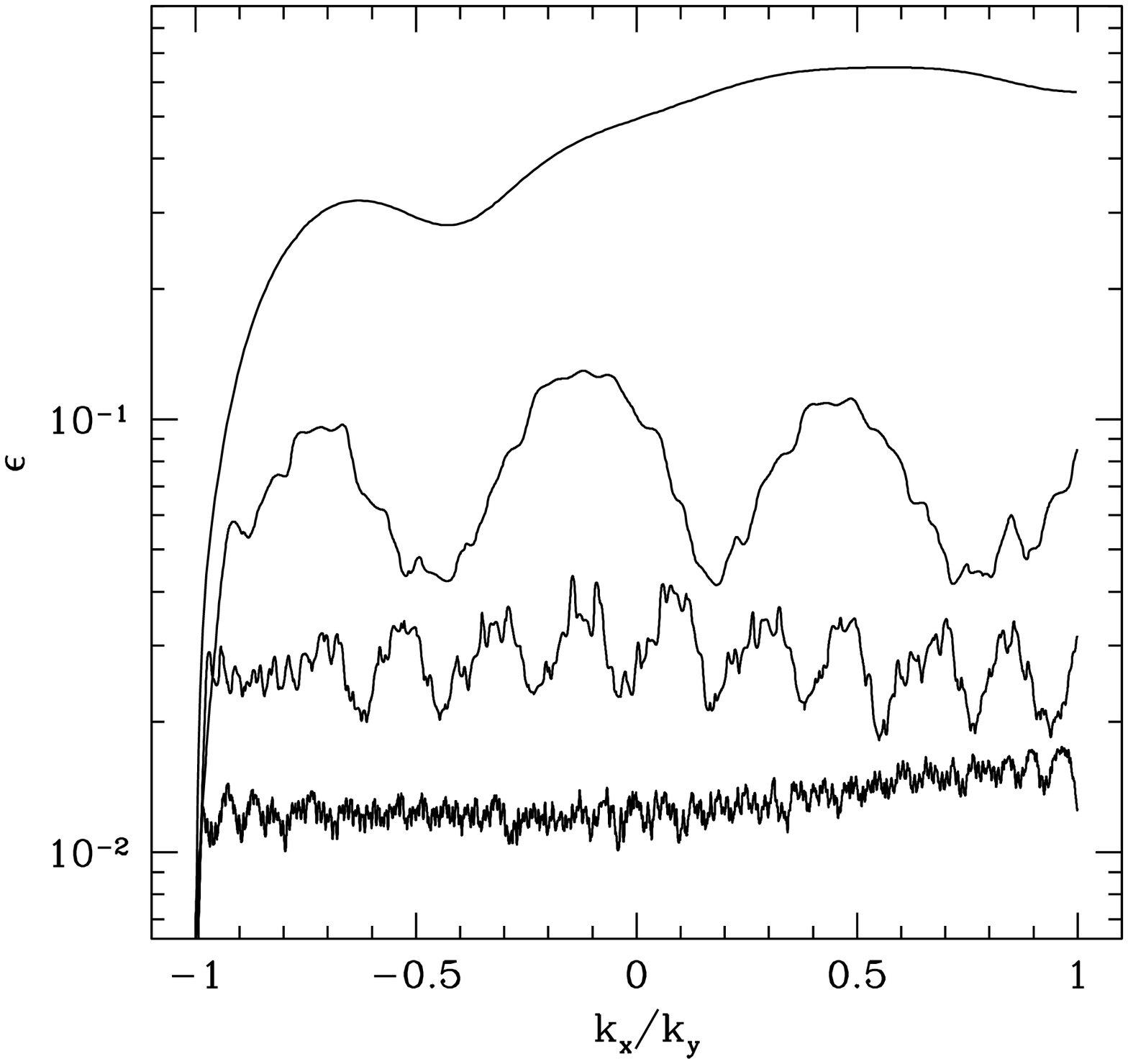}
\caption{
Error in the energy of a slow shwave for a series of numerical integrations with $q = 3/2$ and $c_sk_y/(2\pi\Omega) = c_s k_z/(2\pi\Omega) = 1$ (top), $3$, $10$, $30$ and $100$ (bottom). The scale factor $f$ is defined in expression (\ref{SCALE}). The data have been smoothed for readability.
}
\label{f4}
\end{figure}

Figure~\ref{f4} shows the error in the energy of a slow shwave for a series of numerical integrations with varying $k_z = k_y$ and $q = 3/2$, where again the integration time has been scaled by $f$. The mean field in this case is $v_{Ax} = 0$, $v_{Ay} = c_s$ and $v_{Az} = 0.1c_s$. 

There appear to be two additional regimes in which these solutions break down. Notice from equations (\ref{EVc1}) and (\ref{EVc2}) that the compressive shwaves are singular at $\tilde{\omega} = 0$ , which occurs when $\left(\kdva\right)^2 - v_A^2 k^2 \propto \sin^2 \theta = 0$, i.e., when $\bk$ and $\bv_A$ are parallel. There is a small subset of wavenumbers, therefore, for which the solution breaks down as the angle between $\bk$ and $\bv_A$ becomes small. With $v_{Ax} = 0$, for example, this occurs at $k_x(t) = 0$ and $v_{Az}/v_{Ay} \approx k_z/k_y$. In addition, for wavenumbers satisfying expressions (\ref{KZCRIT}) and (\ref{KXCRIT}), the presence of an unstable incompressive shwave can significantly alter the compressive solutions. There is always coupling between shwaves \citep{gcsl04} since their identity as distinct modes is only accurate to within the errors of the solutions; coupling to an unstable shwave will eventually dominate the evolution of an initially stable shwave.

\subsection{Hydrodynamic Limit}

In the absence of magnetic fields, the amplitude of the compressive shwaves is given by
\be
\delv = \sqrt{\omega} \, \uv{k}
\ee
and
\be
\delr = \frac{\sqrt{\omega}}{c_s},
\ee
where $\omega = c_s k$. In the limit $k_z \rightarrow 0$, these reduce to the WKB solutions for short-wavelength compressive shwaves discussed by \cite{jg05}. 

\section{Shwave Energy and Angular Momentum Transport}\label{SEAMT}

The conservation of wave action is directly related to the evolution of wave energy and angular momentum \citep{bal03}. It is straightforward to show that
\be\label{transport}
\pdv{{\cal E}}{t} = \frac{{\cal E}}{\omega} \pdv{\omega}{t} = q\Omega{\cal T}_{xy}
\ee
for both the incompressive and compressive shwaves, where ${\cal T}_{xy}$ is the phase-averaged shear stress $\delta v_x \delta v_y - \delbx \delby$. Two things are apparent from these relations: 1) The average energy of a shwave grows in proportion to its frequency, and 2) the direction of angular momentum transport associated with a shwave can be determined from the time derivative of its frequency.\footnote{\cite{bal03} gives a comprehensive discussion of wave energy and angular momentum and their relation to action conservation in terms of spatial fluxes of definite frequency modes. There are no net spatial fluxes in a local model such as the shearing box, but one can determine the direction of global transport from the sign of the local shear stress, where a positive (negative) shear stress implies outward (inward) transport (e.g., \citealt{rg92,bh98,bal00}).} The former implies that the energy of a compressive shwave (which has $\omega \propto k$ or $\omega \propto v_A k$ when $v_{Ax} \neq 0$) grows asymptotically (i.e., when it is tightly wound) and is a minimum when it is in an open configuration (i.e., when $k_x = 0$), whereas the energy of an incompressive shwave (which has $\omega \propto 1/k$) peaks at $k_x = 0$ and decays asymptotically. The latter implies that a compressive shwave (which has $\dot{\omega} \propto k_x$) transports angular momentum outward (inward) when it is in a trailing (leading) configuration, whereas an incompressive shwave (which has $\dot{\omega} \propto -k_x$) has the opposite behavior.\footnote{These considerations assume that $\omega$ is real, a necessary condition for the conservation of wave action; magnetorotationally unstable modes are incompressive and yet transport angular momentum outwards.}

\section{Summary and Discussion}\label{section_summ}

To within a multiplicative (dimensional) constant, the solution for incompressive shwaves (\S\ref{section_is}) in an isothermal MHD shear flow is given to first order by:
\bea
\delv = \tilde{\omega}\sqrt{\frac{\omega}{2\tilde{\omega}^2 k^2 -  \kappa^2 k_z^2}}\, \bk \times \left(\bk \times \uv{x} \right. \nonumber \\ \; \left. + i\frac{\tilde{\omega}^2 k^2}{2 \Omega k_z \omega}\uv{x}\right)
\exp\left(-i \int \omega \, dt + i \bk \cdot \bld{x}\right),
\eea
\bea
\delb = -\frac{\tilde{\omega} \, \kdva}{\omega}\sqrt{\frac{\omega}{2\tilde{\omega}^2 k^2 -  \kappa^2 k_z^2}}\, \bk \times \left(\bk \times \uv{x} \right. \nonumber \\ \; \left. + i\frac{2\Omega k_z \omega}{\tilde{\omega}^2} \uv{x}\right)\exp\left(-i \int \omega \, dt + i \bk \cdot \bld{x}\right),
\eea
and
\be
c_s^2 \delr = -\bv_A \cdot \delb - i \frac{2\Omega}{k}\left(\frac{k_x}{k} \uv{y} 
+ [q-1]\frac{k_y}{k}\uv{x}\right) \cdot \delv ,
\ee
where $\omega$ and $\tilde{\omega}$ are given by expressions (\ref{DRi}) and (\ref{TOM}). The incompressive shwaves have $\delv \perp {\delb} \perp \bk$. They are not shear Alfv\`enic, since the effect of the shear is to increase the angle between $\bv_A$ and $\bk$ so that they are asymptotically perpendicular. Their energy peaks at $k_x = 0$ and decreases as $|k_x| \rightarrow \infty$, and they asymptotically transport angular momentum inward. In general, these solutions are accurate for $k_z/k_y \gg 1$ (with errors $\lesssim 30\%$ for $k_z/k_y \lesssim 1$). Within the approximations required for the assumption of incompressibility ($\beta \gg 1$ and $c_s k/\Omega \gg 1$), they are nonlinear solutions to the basic equations \citep{gx94}. For wavevectors $\bk$ initially aligned nearly perpendicular to the mean magnetic field $\bv_A$, as well as for a very weak field ($\sqrt{\beta} \gg c_s k/\Omega \gg 1$), the solution corresponding to the lower branch of the dispersion relation (\ref{DRi}) is no longer uniformly valid. In that case, the solution is given by expressions (\ref{Bxi})-(\ref{vzi}) multiplied by the factor $\exp\left(i\bk \cdot \bld{x}\right)$. 

To within a multiplicative (dimensional) constant, the solution for short-wavelength compressive shwaves (\S\ref{section_cs}) is given to first order by:
\bea
\delv = \frac{\omega k}{\tilde{\omega}}\sqrt{\frac{\omega}{\omega^4 - \left(\kdva\right)^2 c_s^2 k^2}}\left(\frac{\omega^2}{k^2} \bk - \kdva \, \bv_A \right)
\nonumber \\ \times \exp\left(-i \int \omega \, dt + i \bk \cdot \bld{x}\right), \;\;\;\;\;\;\;\;
\eea
\bea
\delb = \frac{\omega^2 k}{\tilde{\omega}}\sqrt{\frac{\omega}{\omega^4 - \left(\kdva\right)^2 c_s^2 k^2}}\left(\bv_A - \frac{\kdva}{k^2} \, \bk \right)
\nonumber \\ \times \exp\left(-i \int \omega \, dt + i \bk \cdot \bld{x}\right),\;\;\;\;\;\;\;\;
\eea
and
\bea
\delr = \frac{\tilde{\omega} k}{c_s}\sqrt{\frac{\omega}{\omega^4 - \left(\kdva\right)^2 c_s^2 k^2}}
\nonumber \\ \times \exp\left(-i \int \omega \, dt + i \bk \cdot \bld{x}\right),
\eea
where here $\omega$ is given by expression (\ref{DRc}). For the compressive shwaves, $\delv$ and $\delb$ lie in the $\bv_A, \bk$ plane, with $\delv \perp {\delb}$ only for $\bv_A \perp \bk$. Their energy is a minimum at $k_x = 0$ and increases as $|k_x| \rightarrow \infty$, and they asymptotically transport angular momentum outward. In general, these solutions are accurate for $H k \gg 1$. The slow-shwave solution, corresponding to the lower branch of the dispersion relation (\ref{DRc}), breaks down for $\beta \gtrsim 1$, whereas the fast-shwave solution is generally valid for arbitrary $\beta$. Both solutions, however, can be compromised by the presence of a transiently-unstable incompressive shwave, and both break down when ${\bv_A \parallel \bk}$. 

The solutions described here are ideal for convergence testing of MHD algorithms that incorporate a background shear flow. They include both super- and sub-thermal magnetic fields as well as compressive and incompressive velocity perturbations and have been verified to be excellent approximations to the full set of linear equations for the bulk of wavenumbers that fit within the disk. Wave action is conserved for both sets of solutions, and is intimately related to the evolution of their energy and angular momentum transport (\S\ref{SEAMT}). The outward (inward) transport of trailing compressive (incompressive) local disturbances is a general result.

The methods employed in this paper can be used to extend any axisymmetric modal analysis in the shearing box to nonaxisymmetry. If the linear equations can be expressed in a form that does not depend explicitly on $k_y$, the nonaxisymmetric dispersion relation is simply given by the axisymmetric dispersion relation with the two-dimensional wavenumber replaced by the three-dimensional one, provided the resulting time-dependent frequency satisfies the adiabatic approximation. If, in addition, the perturbation Lagrangian is known, the slowly-varying amplitude of the nonaxisymmetric modes can be calculated from the wave action of the axisymmetric modes.

One possible astrophysical application is to study the nonlinear interaction of these solutions or their stability to secondary perturbations \citep{gx94}; such a study might shed some light on the saturation of the magnetorotational instability \citep{bh91} in shearing-box simulations \citep{hgb95,sits04}. In addition, the \cite{gs95} theory of MHD turbulence assumes that energy injection at the forcing scale of the turbulence is roughly isotropic. Whereas the mode interactions in the Goldreich-Sridhar (GS) cascade take place on a time scale $(v_A k)^{-1} \ll \Omega^{-1}$, and the bulk of the inertial range should therefore be unaffected by the shear, the incompressive modes considered here with $v_A k \sim \Omega$ have $\bv_A \perp \bk$ for much of their evolution. It is not clear what effect energy injection into modes with $\bv_A \perp \bk$ will have upon the inertial range in the GS cascade (which consists of Alfv\`en wave packets propagating along the field), but it may be that measurements of the turbulent cascade in shearing-box simulations will differ from the isotropic results \citep{cv00,mg01,clv02}.

\acknowledgements

I thank Charles Gammie, Eliot Quataert and Jeremy Goodman for helpful comments, and the referee for a careful reading of the manuscript. I also thank my mother for finding a spelling error. This work was supported by NASA grant NNG05GO22H and the David and Lucile Packard Foundation.

\begin{appendix}

\section{A. Higher-Order Expansion for Incompressive Shwaves}\label{FOIS}

The first-order equations for the incompressive shwaves are
\be
\dot{\wamp{\eta}{0}} - i\omega \wamp{\eta}{1} - i (\kdva) k^2 \wamp{\bx}{1}
+ 2\Omega k_z \wamp{\omega_x}{1} = 0,
\ee
\be
\dot{\wamp{\omega_x}{0}} - i\omega \wamp{\omega_x}{1} - i (\kdva) \wamp{J_x}{1} - (2 - q) \Omega k_z \frac{\wamp{\eta}{1}}{k^2} = 0,
\ee
\be
\dot{\wamp{\bx}{0}} - i\omega \wamp{\bx}{1} - i (\kdva) \frac{\wamp{\eta}{1}}{k^2} = 0,
\ee
and
\be
\dot{\wamp{J_x}{0}} - i\omega \wamp{J_x}{1} - i (\kdva) \wamp{\omega_x}{1} - q\Omega k_z \wamp{\bx}{1} = 0,
\ee
where $\{\delta^{(0)}\}$ are the lowest-order solutions from \S\ref{SIS} and $\{\delta^{(1)}\}$ are higher-order corrections. In the absence of the time derivatives of the zeroth-order amplitudes, the above set of equations is the same eigenvalue problem solved in \S\ref{SIS}. As a result, the combination of the first-order equations results in a term proportional to the dispersion relation (\ref{DRi}) times one of the first-order amplitudes. This term is therefore of a higher order in the expansion and can be neglected, and one ends up with an equation for the time evolution of the zeroth-order amplitudes:
\be
\dwamp{\omega_x}{0} - \frac{\kdva}{\omega} \dwamp{J_x}{0} - i\frac{2\Omega \bv_A \cdot 
\bk k_z}{\tilde{\omega}^2} \dwamp{\bx}{0} + i \frac{\Omega k_z \omega}{\tilde{\omega}^2 k^2} \left(2 - q\frac{\tilde{\omega}^2}{\omega^2}\right) \dwamp{\eta}{0} = 0.
\ee

Expressing all the amplitudes in terms of $\wamp{\bx}{0}$ via the eigenvector relationships (\ref{EVi1}) and (\ref{EVi2}), which can also be written as
\be\label{EVi}
\left(\wamp{\eta}{0}, \wamp{\omega_x}{0}, \wamp{J_x}{0}\right) = \left(-\frac{\omega k^2}{\kdva}, -i\frac{\tilde{\omega}^2 k^2}{2 \Omega k_z \kdva}, i\frac{2 \Omega k_z \omega}{ \tilde{\omega}^2}\right)\wamp{\bx}{0},
\ee
and combining terms gives
\be\label{AMPi}
\left(2\tilde{\omega}^2 k^2 - \kappa^2 k_z^2\right)\dv{}{t}\left(\ln\wamp{\bx}{0}\right) + 
\dv{\left(\tilde{\omega}^2 k^2\right)}{t}
+ \frac{\tilde{\omega}^2 k^2 - \kappa^2 k_z^2}{2\omega}\dv{\omega}{t}
+ \frac{\tilde{\omega}^2}{2\omega}\dv{\left(\omega k^2\right)}{t} = 0.
\ee
Notice that this is a first-order differential equation; all of the degrees of freedom are contained within the dispersion relation (\ref{DRi}).

Using the following relation obtained from differentiation of the dispersion relation (\ref{DRi}):
\be
\left(k_z^2\kappa^2 - 2\tilde{\omega}^2 k^2\right)\dv{\tilde{\omega}^2}{t} = \tilde{\omega}^4\dv{k^2}{t},
\ee
equation (\ref{AMPi}) can be expressed as
\be\label{AMPi2}
\dv{}{t}\left(\ln\wamp{\bx}{0}\right) + 
\frac{1}{2\tilde{\omega}^2 k^2 - \kappa^2 k_z^2} \dv{\left(\tilde{\omega}^2 k^2\right)}{t}
- \frac{1}{\tilde{\omega}}\dv{\tilde{\omega}}{t} + \frac{1}{2\omega}\dv{\omega}{t} = 0,
\ee
the direct integration of which yields expression (\ref{DBXamp}).

\section{B. Higher-Order Expansion for Compressive Shwaves}\label{FOCS}

The first-order equations for the compressive shwaves are
\be
\dwamp{\rho}{0} - i\omega \wamp{\rho}{1} + i \bk \cdot \wamp{\bv}{1} = 0,
\ee
\be
\bk \cdot \dwamp{\bv}{0} - i \omega \bk \cdot \wamp{\bv}{1} + i k^2 \left(c_s^2 \wamp{\rho}{1} + \bv_A \cdot \wamp{\bva}{1}\right) + (2 - q)\Omega k_y \wamp{v_x}{0} - 2\Omega k_x \wamp{v_y}{0} = 0,
\ee
\be
\bv_A \cdot \dwamp{\bv}{0} - i \omega \bv_A \cdot \wamp{\bv}{1} + i \kdva c_s^2 \wamp{\rho}{1} + (2 - q)\Omega v_{Ay} \wamp{v_x}{0} - 2\Omega v_{Ax} \wamp{v_y}{0} = 0,
\ee
and
\be
\bv_A \cdot \dwamp{\bva}{0} - i\omega \bv_A \cdot \wamp{\bva}{1} - i\kdva \bv_A \cdot \wamp{\bv}{1} + iv_A^2 \bk \cdot \wamp{\bv}{1} + q\Omega v_{Ay} \wamp{\bx}{0} = 0,
\ee
where $\{\delta^{(0)}\}$ are the lowest-order solutions from \S\ref{SCS} and $\{\delta^{(1)}\}$ are higher-order corrections. As in Appendix~\ref{FOIS}, these can be combined into a single equation for the time dependence of the zeroth-order amplitudes:
\bea\label{AMPc}
\bk \cdot \dwamp{\bv}{0} 
+ \frac{c_s^2 k^2\tilde{\omega}^2}{\omega^3} \dwamp{\rho}{0}
+ (2 - q)\Omega k_y \uv{x} \cdot \wamp{\bv}{0} - 2\Omega k_x \uv{y} \cdot \wamp{\bv}{0} + \frac{k^2}{\omega} \left(q\Omega v_{Ay} \uv{x} \cdot \wamp{\bva}{0}
+ \bv_A \cdot \dwamp{\bva}{0} \right. \nonumber \\ \; \left. - \frac{\kdva}{\omega}\left[\bv_A \cdot \dwamp{\bv}{0} + (2 - q)\Omega v_{Ay} \uv{x} \cdot \wamp{\bv}{0} - 2\Omega v_{Ax} \uv{y} \cdot \wamp{\bv}{0}\right]\right) = 0.
\eea
Notice that the time dependence of the mean magnetic field (equation [\ref{VA}]) implies
\be
\dv{}{t} \left(\bv_A \cdot \wamp{\bv}{0}\right) = \bv_A \cdot \dwamp{\bv}{0} - q\Omega v_{Ax} \uv{y} \cdot \wamp{\bv}{0},
\ee
with an analogous expression for $\bv_A \cdot \wamp{\bva}{0}$.

Considerable simplification of equation (\ref{AMPc}) results from combining the radial and azimuthal components of the velocity and magnetic field, which are given by the dot product of $\uv{x}$ and $\uv{y}$ with (\ref{EVc1}) and (\ref{EVc2}). In particular, the Coriolis terms cancel, and the terms proportional to $q\Omega$ can be expressed in terms of derivatives of $k$ and $v_A$. Performing this simplification and expressing all of the amplitudes in terms of $\bk \cdot \wamp{\bv}{0}$ via the eigenvector relationships (\ref{EVc1}) and (\ref{EVc2}), rewritten as
\be\label{EVc}
\left(\wamp{\rho}{0}, \bv_A \cdot \wamp{\bv}{0}, \bv_A \cdot \wamp{\bva}{0} \right)   = \left(1,\frac{c_s^2 [\kdva]}{\omega}, \frac{\omega^2 - c_s^2 k^2}{k^2} \right) \frac{\bk \cdot \wamp{\bv}{0}}{\omega},
\ee
one obtains the following equation for the amplitude of the compressive shwaves:
\bea\label{AMPc2}
2\frac{\omega^4 - c_s^2 k^2\left(\kdva\right)^2}{\omega^4}\dv{}{t}\left(\ln \bk \cdot \wamp{\bv}{0}\right)
+ \frac{k^2}{\omega} \dv{}{t}\left(\frac{\omega^2 - c_s^2 k^2}{\omega k^2}\right)
\nonumber \\ \; 
- \frac{\left(\kdva\right)^2 c_s^2 k^2}{\omega^2} \dv{}{t}\left(\frac{1}{\omega^2} \right) 
+ \frac{c_s^2 k^2\tilde{\omega}^2}{\omega^3} \dv{}{t}\left(\frac{1}{\omega}\right)
- \frac{1}{\tilde{\omega}^2}\left(k^2 \dv{v_A^2}{t} + \frac{\omega^2}{k^2}\dv{k^2}{t}\right) = 0.
\eea

Differentiation of the dispersion relation (\ref{DRc}) with respect to $t$ yields the following relationship between the time derivatives of $k$, $v_A$ and $\omega$:
\be
k^2 \dv{v_A^2}{t} + \frac{\omega^2}{k^2}\dv{k^2}{t} = \frac{\omega^4 - \left(\kdva\right)^2 c_s^2 k^2}{\omega^4}\dv{\omega^2}{t}.
\ee
Using this and
\be
\frac{k^2}{\omega^2} \dv{}{t}\left(\frac{\omega^2}{k^2}\right) = \frac{k^2}{\omega^4}\dv{}{t}\left(\frac{\omega^4 - \left[\kdva\right]^2 c_s^2 k^2}{k^2}\right) - \frac{2}{\omega}\dv{\omega}{t},
\ee
and combining terms proportional to $\dot{\omega}$, the amplitude equation (\ref{AMPc2}) simplifies to 
\be\label{AMPc3}
\dv{}{t}\left(\ln \bk \cdot \wamp{\bv}{0}\right)
- \frac{1}{\tilde{\omega}}\dv{\tilde{\omega}}{t}
- \frac{3}{2\omega} \dv{\omega}{t}
+ \frac{k^2}{\omega^4 - \left(\kdva\right)^2 c_s^2 k^2}\dv{}{t}\left(\frac{\omega^4 - \left[\kdva\right]^2 c_s^2 k^2}{2k^2}\right) = 0.
\ee
Direct integration of the above expression gives
\be
\bk \cdot \wamp{\bv}{0}
\propto \omega\tilde{\omega}k\sqrt{\frac{\omega}{\omega^4 - c_s^2 k^2 \left(\kdva\right)^2}},
\ee
which, along with $\bk \cdot \wamp{\bv}{0} = \omega \wamp{\rho}{0}$, is equivalent to expression (\ref{DRamp}).

\section{C. Shwave Energy}\label{section_se}

The second-order Lagrangian\footnote{Second-order here refers to an expansion in powers of a small-amplitude displacement.} for a rotating, isothermal MHD shear flow (with $\rho_0 \equiv 1$) is given by \citep{dew70,hay70,ngg87}
\bea\label{Lag2}
L_2 = \frac{1}{2}\left(\left|\dv{\bxi}{t}\right|^2 + 2\left[\bO \times \bxi\right] \cdot \dv{\bxi}{t} + 2q\Omega \xi_x^2 - c_s^2 \pdv{\xi_i}{x_j} \pdv{\xi_j}{x_i} \right. \nonumber \\ \left. \; - \frac{v_A^2}{2} \left[\left\{\bnabla \cdot \bxi\right\}^2 + \pdv{\xi_i}{x_j} \pdv{\xi_j}{x_i}\right] + 2\left[\bnabla \cdot \bxi\right] \bv_A \cdot \left[\bv_A \cdot \bnabla \bxi\right] - \left[\bv_A \cdot \bnabla \bxi\right]^2\right),
\eea
Variation of this perturbation Lagrangian yields the linearized equations of motion (\ref{L1})-(\ref{L4}). Inserting the real part of the perturbations (including the exponential factor) into expression (\ref{Lag2}) and averaging over azimuth gives
\bea
{\cal L} \equiv k_y \int_{0}^{2\pi} L_2 \, dy = \frac{1}{4}\left(\omega^2\left| \xi \right|^2 + 4\omega \bO \cdot \left[\bxi_r \times \bxi_i\right] + 2q\Omega^2\left| \xi_x \right|^2 - \left[c_s^2 + v_A^2\right] \left| \bk \cdot \bxi \right|^2 \right. \nonumber \\ \; \left. + \; 2 \, \left[\kdva\right] \left[\bk \cdot \bxi_i \bv_A \cdot \bxi_i + \bk \cdot \bxi_r \bv_A \cdot \bxi_r\right] - \left[\kdva\right]^2 \left| \xi \right|^2\right),
\eea
where $|\xi|^2 = \xi_r^2 + \xi_i^2$, and $\bxi_r$ and $\bxi_i$ are the real and imaginary parts of the fluid displacements (without the exponential factor). Noting that the fluid displacements are related to the density and magnetic field perturbations by
\be
\delr = -i \bk \cdot \bxi
\ee
and
\be
\delb = i(\kdva \, \bxi - \bk \cdot \bxi\,\bv_A)
\ee
via the linearized continuity and induction equations, the averaged perturbation Lagrangian can be expressed in simpler form as
\be
{\cal L} = \frac{1}{4}\left(\omega^2\left| \xi \right|^2 + 4\omega \bO \cdot \left[\bxi_r \times \bxi_i\right] + 2q\Omega^2\left| \xi_x \right|^2 - c_s^2 \left| \delr \right|^2 - \left|\delta v_A \right|^2\right).
\ee

The average shwave energy as defined in expression (\ref{Edef}) is thus
\be\label{SE}
{\cal E} = \frac{1}{4}\left(\omega^2\left| \xi \right|^2 - 2q\Omega^2\left| \xi_x \right|^2 + c_s^2 \left| \delr \right|^2 + \left|\delta v_A \right|^2\right).
\ee

\end{appendix}

\newpage

\end{document}